\newif\ifsubmission
\submissionfalse

\ifsubmission

\else

\documentclass[webversion]{usenix-2019}

\usepackage{url}
\usepackage{cite}
\usepackage{graphicx}
\usepackage{graphics}
\usepackage{epstopdf}
\usepackage{subfigure}
\usepackage{multirow}
\usepackage{listings}
\usepackage{pifont}
\usepackage{tabularx}
\usepackage{hhline}
\usepackage{paralist}
\usepackage{enumitem}
\usepackage{xspace}
\usepackage{amssymb}
\usepackage[subtle]{savetrees}
\usepackage[T1]{fontenc}
\usepackage{fancyvrb}
\usepackage{xcolor}             
\usepackage[]{hyperref}         
\hypersetup{                    
  colorlinks,
  linkcolor={green!80!black},
  citecolor={red!70!black},
  urlcolor={blue!70!black}
}


\definecolor{darkgreen}{rgb}{0,0.5,0}
\definecolor{purple}{rgb}{0.75,0,0.75}
\definecolor{brown}{rgb}{0.65,0.16,0.16}
\definecolor{darkslateblue}{rgb}{0.28, 0.24, 0.55}
\definecolor{orange}{rgb}{1.0, 0.647, 0}

\newcount\notenum
\newcommand{\fslnote}[3]        {\global\advance\notenum by 1\textsl{\bf\color{#2}$\blacksquare$[#1\the\notenum: #3]}}

\newcommand{\ie}                {\emph{i.e.},\xspace}
\newcommand{\eg}                {\emph{e.g.},\xspace}
\newcommand{\etal}              {\emph{et~al.}\xspace}

\newcolumntype{C}{>{\centering\arraybackslash}X}

\newcommand{\capfont}                   \small

\newcommand{\smallitemsep}              {\setlength{\itemsep}{-0.5ex}}

\newcommand{\CircledOne}        {\ding{172}\xspace}
\newcommand{\CircledTwo}        {\ding{173}\xspace}
\newcommand{\CircledThree}      {\ding{174}\xspace}
\newcommand{\CircledFour}       {\ding{175}\xspace}
\newcommand{\CircledFive}       {\ding{176}\xspace}

\newcommand{\stt}[1]    {{\tt\small #1}}

\newcommand{\kml}{KML\xspace}

\begin{document}

\title{KML: Using Machine Learning to Improve Storage Systems\\
}
\author{Ibrahim Umit Akgun, Ali Selman Aydin, Andrew Burford, Michael McNeill, Michael Arkhangelskiy,\\
Aadil Shaikh, Lukas Velikov, and Erez Zadok\\
\authaddr{Stony Brook University}
}
\maketitle
\pagestyle{plain}
\begin{abstract}

Operating systems include many heuristic algorithms designed to
improve overall storage performance and throughput.
Because such heuristics cannot work well for all conditions and
workloads, system designers resorted to exposing numerous tunable
parameters to users---thus burdening users with continually
optimizing their own storage systems and applications.
Storage systems are usually responsible for most latency in I/O-heavy
applications, so even a small latency improvement can be
significant.
Machine learning (ML) techniques promise to learn patterns, generalize
from them, and enable optimal solutions that adapt to changing
workloads.
We propose that ML solutions become a first-class component in OSs and
replace manual heuristics to optimize storage systems dynamically.
In this paper, we describe our proposed ML architecture, called \kml.
We developed a prototype \kml architecture and applied it to two
case studies: optimizing readahead and NFS read-size values.
%
%
Our experiments show that \kml consumes less than 4KB of dynamic
kernel memory, has a CPU overhead smaller than 0.2\%, and yet can
learn patterns and improve I/O throughput by as much as 2.3$\times$ and
15$\times$ for two case studies---even for complex,
never-seen-before, concurrently running mixed workloads on different
storage devices.

\end{abstract}


\section{Introduction}
\label{sec:introduction}

Computer hardware, software, storage, and workloads are constantly
changing.  Storage performance
heavily depends on workloads and the precise system
configuration~\cite{fast17instability, fast10fsgreen}.  Storage
systems and OSs include many parameters that can affect overall
performance~\cite{atc18auto-tuning, fast20carver, yadgar2021ssd}.
Yet, users often do not have the time or expertise to tune these
parameters.  Worse, the storage and OS communities are fairly
conservative and resist making significant changes to
systems to prevent instability or data loss.  Thus, many techniques
currently used were historically developed with human intuition
after studying a few workloads; but such techniques cannot easily
adapt to ever-changing workloads and system diversities.

For example, readahead values, while tunable, are often fixed and left
at their defaults.   Correctly setting them is important and difficult
when workloads change: too little readahead wastes potential
throughput and too much pollutes caches---both hurting
performance.
Some OSs let users pass hints (\eg{} \stt{fadvise}, \stt{madvise}) to
help recognize files that will be used sequentially or
randomly, but these often fail to find optimal values for complex,
mixed, or changing workloads.
We experimented with a variety of modern workloads and mainy different
values of readahead: in prior work, we confirmed that no single
readahead value is optimal for all
workloads~\cite{hotstorage21kml}.
Another example of tunable parameters in the network storage settings is the
default read-size (\stt{rsize}) parameter in NFS: if set too
small or large, performance suffers.

Machine Learning (ML) techniques can address this complex relationship
between workloads and tunable parameters by observing actual
behavior and adapting on-the-fly, and hence may be more promising than
fixed heuristics.
ML techniques were recently used to predict index
structures in KV stores~\cite{kraska2018case, dai2020wisckey}, for
database query optimization~\cite{sagedbkraska}, improved
caching~\cite{subedi2018stacker}, cache eviction
policies~\cite{vietri-hotstorage18-lecar}, I/O
scheduling~\cite{hao2020linnos}, and more.

In this paper, we describe our ML approach to improve storage
performance by dynamically adapting to changing I/O workloads.
We designed and developed a versatile, low-overhead, light-weight
system called \emph{\kml}, for conducting ML training and prediction
for storage systems.  \kml defines generic ML APIs that can be
used for a variety of subsystems; we currently support several deep
neural networks and decision tree models.
We designed \kml to be embeddable inside an OS or the critical path of
the storage system: \kml imposes low CPU and memory overheads.
\kml can run synchronously or asynchronously, giving users the ability
to trade-off prediction accuracy vs.\ overhead.

Developing and tuning ML-based applications can be its own challenge.
Therefore, we designed \kml to run identically in user- or
kernel-level.  Users can develop and debug ML solutions easily in the user
level, then upload the same model to run identically in the kernel.

We demonstrate \kml's usefulness with two case studies:
(i) adapting readahead values dynamically and (ii) setting NFS
\stt{rsize} values automatically.  In both cases, we aim to adapt these
values within one second
under changing and even mixed workloads.
Overall, our approach to storage systems optimization using ML is a
continuous \emph{observe-and-tune} paradigm.

This paper makes five contributions:
\begin{enumerate}
\smallitemsep
\item We show that lightweight ML can indeed become a first-class
  citizen inside storage systems and OSs;
\item We offer flexibility through synchronous or asynchronous
  training and the ability to offload training to the user-level;
\item We introduce the idea of generic ML APIs that can be expanded to
  support additional and future ML techniques;
\item We apply \kml to two important optimization problems (readahead
  and NFS \stt{rsize} values);
and
\item We evaluate our solutions using multiple, complex, and even
  mixed workloads, as well as two different storage devices.  We
  demonstrate throughput improvements up to 2.3$\times$ for
  readhead and up to 15$\times$ for \stt{rsize}.  We show that ML
  models trained on a few workloads can generalize and optimize
  throughput for never-before-seen workloads or devices.  And finally,
  we show that \kml has small CPU overheads ($<0.2\%$) and dynamic memory
  footprint (4KB),
  well worth the overall I/O improvements.
\end{enumerate}

Next,
Section~\ref{sec:kml-arch} describes \kml's design.
Section~\ref{sec:use-cases} describes our two use cases (readahead and
NFS \stt{rsize}).
Detailed evaluation of \kml and two use cases are in
Section~\ref{sec:evaluation}.
We survey related work in Section~\ref{sec:related} and conclude in
Section~\ref{sec:conclusion}.


\section{\kml's Architecture}
\label{sec:kml-arch}

Modern ML libraries are often general-purpose, rely on many large
third-party libraries (\eg in C++ or Python), and designed to process
lots data using massive processing power (\eg GPU clusters).
Porting such ML systems to an OS kernel would be impractical, because
an OS is a highly constrained and unforgiving environment.
Thus, we chose to develop an ML framework from scratch---designed for
low-overhead, light-weight, and highly tailored to OSs and storage
systems and developers.

\paragraph{\kml high-level design choices}
%


\begin{figure}[t!]
  \centering
  \includegraphics[width=0.9\columnwidth]{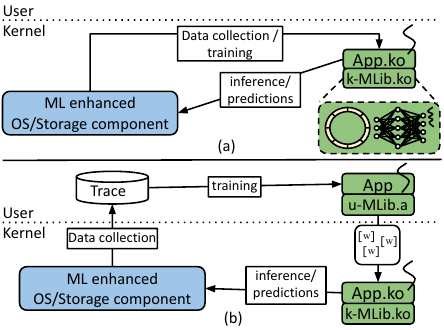}
  \caption{Two different operational modes that we built to
    achieve a high efficiency ML framework for tuning OS-level storage
    systems: (a) kernel space training and inference and (b) offline user
    space training and kernel space inference.}
  \label{fig:arch-comparison}
\end{figure}


%
Figure~\ref{fig:arch-comparison} demonstrates two different operating
modes that we built.
\kml supports (a) in-kernel training and
inference and (b) user space offline training and in-kernel inference.
Once a model is built in user space, it can be loaded into the kernel
as is.
\kml has a highly modular design: the core ML code base is shared by
both user and kernel space.
Operation mode (a) is designed for performance and accuracy,
especially under high-I/O rates, because collecting and copying lots
of I/O event data out of the kernel imposes high overheads.
Operation mode (b) is designed to simplify ML model development for
OS/storage developers.  Users can develop and test an ML model design
more easily in user space, testing different features, ML
architectures, and hyper-parameters to reach a stable and accurate
model.

\subsection{Design Overview}
\label{sec:kml-design-overview}

\paragraph{Easy to develop and extend}
In Figure~\ref{fig:arch-comparison}(b), \kml is compiled and linked
with an application for both kernel and user space.  \stt{u-MLib.a}
and \stt{k-MLib.ko} are built using the same \kml source code.  We
developed a wrapper layer for the \kml development API: \kml's core
code is uniform across both user and kernel APIs.  This identical
abstraction speeds up development, eases debugging, and facilitates
extensibility (see Section~\ref{sec:kml-modular-design}).
Nevertheless, we recognize that while we aim to make ML-based
solutions easier to use, developers still require a good understanding
of OS and storage system internals.

\paragraph{Low overhead}
To make ML approaches practical for storage systems, they must have
low computational and memory overheads.  ML solutions have three
phases that consume much memory/CPU resources: (i) inference (\ie
prediction), (ii) training, and (iii) data processing \&
normalization.
We support asynchronous training and inference capabilities to reduce
interference on the data path; \kml also uses efficient communications
between the data collection and model training \& inference
components, to help scalability and stability of ML-based designs.
To reduce the data collection overheads, developers can facilitate
subsampling techniques that are provided in \kml.
We detail our design choices to reduce these overheads
in Section~\ref{sec:kml-overheads}.


\subsection{Fundamentals of Core ML library}
\label{sec:kml-core-ml}

\kml provides primitives for building and extending ML models.
This involves building algorithms for training ML models (\eg
back-propagation, decision-tree induction) and building the mathematical
functions needed to implement them.  The library design allows for
seamless extensibility of library functionality.  Additionally, our ML
functionality is easily debugged in user space as it uses identical
code and APIs in kernel space.

\paragraph{Mathematical and matrix operations}
Most ML algorithms rely heavily on basic
mathematical functions and matrix algebra.  For example, a neural
network classifier uses functions such as matrix
multiplication/addition, \stt{softmax}, and exponentiation.  Hence, we
implemented kernel versions of such common ML functions using well known
approximation algorithms.

\paragraph{Layer and loss-function implementations}
One can think of a neural network as a composition of layers and one
or more loss functions.  Many of these building blocks are used across
many different neural network architectures.  Layers like a fully
connected layer, ReLU~\cite{nair2010rectified}, or sigmoid are
essential building blocks of many neural networks; loss functions are
also fairly common across many applications.  Both
layers and loss functions implement two main functionalities, one
during the inference (forward) phase and another during the
back-propagation (training) phase.
We implemented these common components and their forward and back-propagation functionality
from scratch in \kml: layer/loss functions, data structures related to the layer/loss, etc.

\paragraph{Inference and training}
When stacked together, the elements of a conventional neural network
can form a DAG.  Thus, a neural network inference means traversing the
DAG starting from the initial node(s) (where the inputs are provided),
toward the resulting nodes (where the neural network output is
produced).  \kml implements a standard training method used in neural
networks---back-propagation~\cite{rumelhart1986learning}.
\kml also includes Stochastic Gradient Descent (SGD) which uses the gradients
computed using back-propagation to optimize the neural network weights.


\subsection{\kml's Modular Design}
\label{sec:kml-modular-design}

We now elaborate on \kml's operation modes: (i) in-kernel training and
inference (see Figure~\ref{fig:arch-comparison}(a)), and (ii) user
space training and in-kernel inference (see
Figure~\ref{fig:arch-comparison}(b)).

\paragraph{Training in kernel space}
%

\begin{figure}[t!]
  \centering
  \includegraphics[width=1.0\columnwidth]{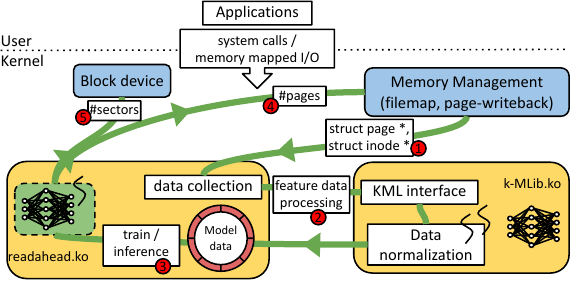}
  \caption{\kml kernel space training and inferencing
    architecture.}
  \label{fig:arch-online}
\end{figure}


%
We use the readahead use case to describe how \kml works in kernel
training and inference mode.  Figure~\ref{fig:arch-online} shows \kml's
framework (\stt{k-MLib.ko}), a \kml application (\stt{readahead.ko}), and
target storage components (Block device and Memory Management
subsystems).  The yellow background denotes \kml related components.  The
blue background depicts the target storage components, which are
specific to the readahead case.  The green line represents execution and
data flow.
Numbered boxes refer to transitions happening between the
components.

As we mentioned in Section~\ref{sec:introduction}, we designed our use
cases based on a continuous \emph{observe-and-tune} principle.  In its first
stage, the readahead module observes and collects data.  Since
our target component is the memory management (\eg page cache)
system, the readahead
module starts collecting data from this component
(Figure~\ref{fig:arch-online}~\CircledOne).
The readahead module then extracts features and transfers them to the
\kml framework to be normalized (Figure~\ref{fig:arch-online}~\CircledTwo).
After the data processing and normalization stage is done, if the
readahead module is operating in training mode, it trains on the
normalized data, and the execution flow ends here.  However, if the
readahead module is operating in inference mode, it feeds the
normalized data to the readahead neural network model and tunes the
target components based on the model's prediction
(Figure~\ref{fig:arch-online}~\CircledThree).

How a \kml application optimizes a target component depends on the
problem and its solution.  Here, the readahead module updates
readahead sizes on a per-file basis (Figure~\ref{fig:arch-online}~\CircledFour)
or a per-device basis (Figure~\ref{fig:arch-online}~\CircledFive).
When the readahead module is inferencing, execution flow forms a
\emph{closed-circuit}.
After the readahead module changes readahead sizes, OS memory state
changes; thereafter, new inputs go to the readahead neural network
model, leading to updated predictions.
Therefore, ML is particularly suitable to solve problems that require
an ongoing cycle of observing and tuning.
%

In the ML ecosystem, data collection is a crucial part.  One reason we
offer kernel training is to train on data collected with a high
sampling rate.  Tracing OSs and storage systems with high accuracy and
sampling rates is challenging~\cite{systor20reanimator}.
Nevertheless, tracing tools like LTTng~\cite{lttng-web} can bring
overhead down to as little as 5\%.
Additionally, traces may still be inaccurate due to data loss.  LTTng
collects trace data in shared user/kernel lockless circular buffers;
under heavy sampling loads, some trace events can be dropped if
LTTng's user-level processing threads do not consume the samples fast
enough.
However, operating in kernel space gives \kml
more control over thread scheduling to reduce loss of sampled events.
Since our use cases may require high sampling rates for I/O events,
placing data processing and normalization in user space would lose too
much valuable data than in the kernel.
Still, we believe a user-kernel co-operated design may be beneficial
in some cases (part of our future work).

\paragraph{Training in user space}
Building ML solutions is an iterative process.
To find the essential features and build accurate models,
we need to run multiple data analyses, train, then test an ML model with
different architectures and hyper-parameters.
To speed up model development and debugging, \kml offers
offline user-space training and kernel inferencing mode (see
Figure~\ref{fig:arch-comparison}(b)).  As \kml's user- and
kernel-space libraries use the same APIs and code base, models trained
in user space can be loaded into the kernel as is.
%

\begin{figure}[t!]
  \centering
  \includegraphics[width=1.0\columnwidth]{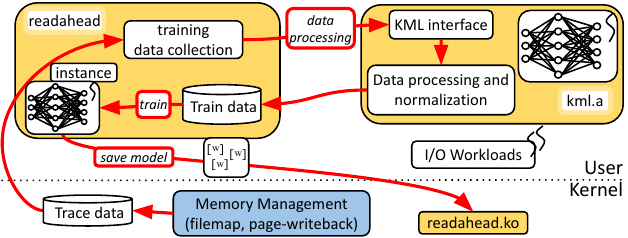}
  \caption{\kml user-space training \& kernel-space inference
    architecture.}
  \label{fig:arch-offline}
\end{figure}


%
Figure~\ref{fig:arch-offline} shows how the readahead model works in
operation mode.
Components highlighted in yellow represent \kml-specific
implementations.  The red arrows denote the offline data collection
and training paths.

We started by collecting training data using in-kernel tracing of the
target storage components~\cite{systor20reanimator}.  Next was
feature-extraction; this is where user-space training was useful,
because we could run various analyses, test different features, and
implement many data-normalization techniques without re-running I/O
experiments.
After we finalized the feature selection, we trained and tested the
readahead ML model in user space, varying several hyper-parameters; we
used Tune~\cite{liaw2018tune} to optimize our hyper-parameters.
When the readahead ML model was ready for real-time testing, the only
remaining step was to save the trained model to a \kml-specific file
and load it into the readahead kernel module.
\kml APIs facilitate all the functionality necessary for building,
training, saving, and deploying ML models in-kernel.

To ensure identical kernel and user APIs, we use wrappers to abstract
external functionality.
\kml's development API provides 30 functions that fall into five
categories:
\begin{inparaenum}[(i)]
\item memory management,
\item threading,
\item logging,
\item atomic operations,
and
\item file operations.
\end{inparaenum}
For example, we have a simple wrapper called \stt{kml\_malloc}
that calls \stt{malloc} in user-level and \stt{kmalloc} in kernel
space.
For brevity, API details and prototypes are omitted, but part of our
released code (see Section~\ref{sec:kml-implementation}).


\subsection{Computational \& Memory Overheads}
\label{sec:kml-overheads}

OSs and storage systems are susceptible to performance degradation and
increased latency if computational and memory resources are not
carefully managed.  Therefore, we designed \kml with efficient CPU and
memory usage in mind.  There is often a positive correlation between
the computational and memory footprint of an ML model and its training
and inference accuracy.  Hence, \kml is highly configurable, letting
users trade-off overheads vs.\ prediction accuracy to best suit the
problem at hand.

\paragraph{Reducing computational overheads}
Matrix manipulation is a computationally intensive ML building block
that relies on floating-point (FP) operations.  OSs often disable the
floating-point unit (FPU) in the kernel to reduce context-switching
overheads.  To address this, we considered three approaches:
\begin{inparaenum}[(1)]
\item quantization,
\item fixed-point representations,
and
\item temporarily enabling the FPU unit in kernel space.
\end{inparaenum}
Quantization provides compact representation, allows developers to
compute matrix manipulation operations, and does not require an
FPU~\cite{choi2019accurate, gupta2015deep, desa2018understanding,
  de2018high, hao2020linnos}.  Quantization can help reduce
computational and memory overheads, but it reduces
accuracy~\cite{hubara2017quantized}.
Fixed-point representation computes FP operations using integer
registers.  Since all FP operations are emulated, integration of
fixed-point representation is fairly easy and even faster in
certain cases~\cite{chen2020mlloadbalancing, lin2016fixed}.  However,
fixed-point representation works within fixed ranges which can result in
numerical instability~\cite{lai2017deep}.
Since both accuracy and stability are vital \kml design goals, we
chose a third alternative: \kml temporarily enable the FPU in the
Linux kernel using \stt{kernel\_fpu\_begin} and
\stt{kernel\_fpu\_end}.  To avoid context-switch overheads, we
minimize the number of code blocks that use FPs and keep these blocks
small.

\paragraph{Reducing memory overheads}
Three factors affect \kml's dynamic memory consumption:
\begin{inparaenum}[(1)]
\item ML model-specific data,
\item \kml's internal memory allocations at training and
  inference,
and
\item data collection for both training and inference.
\end{inparaenum}
ML model-specific data and \kml's internal memory usage depend on the
number of layers, layer sizes, and layer types.
\kml uses dynamic memory allocation for all internal usage purposes
(\eg layer gradients); this helps reduce interference and memory
pressure.
\kml gathers input
data in a lock-free circular buffer; then, an \emph{asynchronous
  training thread} trains on gathered data.  When collecting data with
a high sampling rate, the size of the lock-free circular buffer is
important to the ML model's performance and accuracy.  Users need to
configure the size of the circular buffer to account for the data
sampling rate such that the asynchronous training thread can catch up
with processing.  If the size of the circular buffer is misconfigured,
\kml may lose useful training data, which can reduce the resulting ML
model's accuracy.

\paragraph{Operating under resource-constrained conditions}
\kml exposes a memory allocation and \emph{reservation} API for ML
internals.  The primary motivation behind \kml's memory reservation
capabilities is to ensure predictable performance and accuracy, even
under memory pressure.  This allows \kml to operate without worry of
memory allocation lagging or failing, which would hurt performance and
accuracy.

\paragraph{Data processing \& asynchronous training}
To make ML solutions generalizable, data normalization is often
utilized.   \kml supports data normalization
functionalities such as moving average, standard deviation, and
Z-score calculation.  However, data normalization often requires heavy
FP computation.
Thus, \kml supports offloading training, inference, and data
normalization to a separate \emph{asynchronous thread}---away from the
data path itself.  This thread communicates with other \kml components
(\eg data collection) using a lock-free circular buffer.  By default,
we let Linux schedule this kthread as needed; \kml also supports
pinning the kthread to a CPU core, to ensure it gets higher scheduling
priority when high sampling rates are required.

Subsampling is another viable solution to reduce data collection
overheads, which \kml supports.  However, subsampling can reduce
prediction accuracy, so care is needed to select a suitable sampling
rate.  In Section~\ref{sec:eval:kml} we evaluate the impact of
subsampling windows on overheads, prediction accuracy, and overall I/O
performance.


\subsection{Stability \& Explainability}
\label{sec:kml-stability}

Both the training and inference phases for ML solutions can be
computationally intensive.  Except for model initialization and saving
models to files, \kml APIs involve no other I/Os.
\kml's impact on the stability of storage performance is limited to
memory-allocation latency and concurrency.
Memory allocations in both user and kernel space can use locking
mechanisms, which could incur unexpected latencies.  To minimize these
problems, \kml allocates memory only in the asynchronous training
thread.
\kml uses a lock-free circular buffer for data communication and
reserves 512 bytes of additional memory
to further ensure stability under memory-pressure conditions.
Lastly, we applied standard k-fold cross validation techniques to ensure the
stability of our ML solutions.

ML solutions can suffer from unexpected behavior and are harder to
explain.  Conversely, traditional heuristics have well-defined
behaviors often expressed as closed-form formulas.
An ML algorithm may behave erratically when used in new, unforeseen settings,
which could hurt system performance where ML is deployed.
This type of issue is difficult to troubleshoot due to the
long-standing explainability problems that affect ML
models~\cite{agarwal2020neural}.
\kml currently supports two ML models: neural networks and
decision trees.
Decision tree predictions are more
explainable because they are represented
as a tree of successive \textsc{if-then} statements, bisecting the
range of the features considered.
Deep neural networks, however, are more challenging to
explain and verify.  Nevertheless, recent work focuses on
explainability in ML~\cite{jeyakumar2020can, samek2020toward,
  ras2018explanation, agarwal2020neural}.
While we plan to improve \kml model stability  using
feedback-based control algorithms in the future,
we currently focus on demonstrating that ML \emph{can} tune
storage system parameters \emph{better} than existing heuristics.
%


\subsection{Implementation}
\label{sec:kml-implementation}

\kml contains 12,213 lines of C/C++ code (LoC).
\kml's core ML part has 5,539 LoC, which can be compiled in both user
and kernel space.
Our readahead neural network model code is nearly 1K LoC long: 486 LoC
for collecting data, initializing the model, creating an inference
thread, and changing block-level and file-level readahead sizes;
and another 351 LoC for model definition, data processing, and normalization.
Our NFS neural network model also includes nearly 1K LoC: 435 LoC for
data collection, model initialization, and running inference to
predict workload type; and 338 LoC for creating the model and
manipulating data.

\emph{All of our code has been released on github (\url{https://github.com/sbu-fsl/kernel-ml}~\cite{hotstorage21kml}),
which includes examples, sample data, models, and full API documentation (all 30 methods).}



\section{Use Cases}
\label{sec:use-cases}

We now detail our two use cases: (1) readahead neural network and
decision tree models and (2) NFS neural network model.  We describe the
following for each:
\begin{inparaenum}[(i)]
\item problem definition,
\item data collection for training,
\item data preprocessing and feature extraction,
and
\item building the ML model.
\end{inparaenum}

\subsection{Use Case: Readahead}
\label{sec:usecase-readahead}

\paragraph{Problem definition}
Readahead is a technique to prefetch an additional amount of storage
data into the OS caches in anticipation of its use in the near term.
Determining how much to read ahead has always been challenging: too
little readahead necessitates more disk reads later and too much readahead
pollutes caches with useless data---both hurt performance.
The readahead value is a typical example of a storage system
parameter: while tunable, it is often fixed and left at its default.
Some OSs let users pass hints via \stt{fadvise} and \stt{madvise} to
help the OS recognize files that will be used purely sequentially or
randomly, but these often fail to find optimal values for varied,
mixed, or changing workloads.
Next, we detail our readahead neural network design
(following Figure~\ref{fig:arch-offline}).  Our goal is to predict
optimal readahead sizes while running under dynamic I/O workloads.

\paragraph{Studying the problem}
We experimented with running 4 different RocksDB~\cite{rocksdb}
benchmarks, each with 20 different readahead sizes (8--1024), and
attempted to determine the readahead sizes that yield the best
performance (in ops/sec) for each workload.  This became our training
data, which can help predict readahead values for \emph{other}
workloads and environments.  This investigation revealed that each
workload has a unique behavior and requires a different readahead size
to reach optimal performance.
We further investigated the correlations between file access patterns,
RocksDB workload labels, and performance.  This helped us determine
the information and features needed to build a good model, as
described below.

\paragraph{Data collection}
We used LTTng~\cite{lttng-web} to collect trace data, which we then
used for finding useful features for the readahead problem.  We
captured most page cache tracepoints~\cite{linux-tracepoints} (\eg{}
\stt{add\_to\_page\_cache}, \stt{writeback\_dirty\_page}).
We collected and processed over 20GB
of traces by running multiple
10-minute RocksDB benchmarks on an NVMe-SSD device.
Ten minutes was sufficient for RocksDB to reach a steady state.
After examining these traces, we selected a set of candidate features
based on our domain expertise.
We then picked the features of interest
and decided where to call hook functions which are responsible for
gathering necessary information (\eg{} \stt{struct page}) for
inference.
Our hook functions provide three important raw values:
\begin{inparaenum}[(1)]
\item time difference from the beginning of execution,
\item \stt{inode} number,
and
\item page offsets of the files that were accessed in locations where
  the hooks were called.
\end{inparaenum}

\paragraph{Data preprocessing \& normalization}
We summarize the input data at one-second intervals to ensure we can
quickly adapt to changing I/O workloads while ensuring stability under
short-term activity spikes.
Based on our domain expertise, and through model experimentation, we
selected the following five features for our model:
the number of transactions taking place each second,
the calculated cumulative moving mean
and the cumulative moving standard deviation of page offsets,
the mean absolute page offset differences
for successive transactions, and
the \stt{inode} number (to ensure we process only RocksDB file
accesses).
Before we fed these features to our readahead neural network, we
applied Z-score normalization to each feature.

\paragraph{Why we chose machine learning for this use case}
\begin{figure}[t!]
  \centering
  \includegraphics[width=1.0\columnwidth]{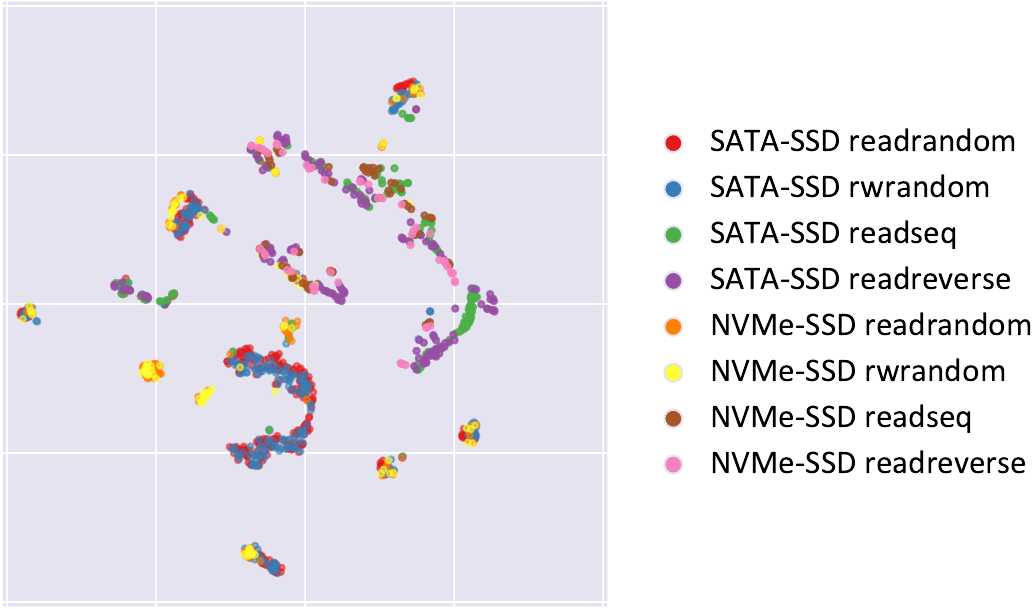}
  \caption{t-SNE visualization of readahead normalized features that
    are generated from both NVMe-SSD and SATA-SSD traces.  Axes are
    intentionally omitted because the dimensions are generated by
    t-SNE and do not represent any specific data.}
  \label{fig:tsne}
\end{figure}


%
After studying the readahead problem, we wanted to explore whether
machine learning would be suitable for solving this problem or whether
more traditional heuristics could still work.  Therefore, while
extracting features from collected traces, we visualized the features
to investigate what type of patterns and clusters the data has.
Figure~\ref{fig:tsne} shows a t-SNE~\cite{van2008visualizing}
visualization of normalized features that are generated from both
NVMe-SSD and SATA-SSD traces.
t-SNE is a visualization technique that applies dimension reduction
and is often used for representing high-dimensional data and looking
for clusters.
We can observe that sequential and random workloads are somewhat
separated; alas, data points from the same workload type are
distributed over multiple clusters, overlapping clusters of other
types.
Worse, even random workloads' clusters overlap with some sequential
workloads' clusters, because RocksDB's warm-up phases involve mostly
sequential accesses---another source of dynamism.
All these findings strongly suggest that workload classification for
the readahead problem would be fairly challenging using traditional
heuristics.
Hence, we felt motivated to explore ML solutions to solving the
readahead problem.

\paragraph{Building neural network model}
We modeled the readahead problem as a classification problem and
designed a neural network with three linear layers (with hidden layer
sizes of 5 and 15), using sigmoid non-linearities in between layers,
and with a cross-entropy loss method as the loss function.  We used an
SGD optimizer~\cite{robbins1951stochastic, kiefer1952stochastic}, and
set a learning rate of 0.01 and a momentum of 0.99 after trying
different values; all these values are common in the
literature~\cite{bengio2012practical}.
We also used Tune~\cite{liaw2018tune} to optimize the
learning rate and momentum.
Our readahead neural network trains on the aforementioned input data
and predicts the workload type.  We trained on the following four
types of RocksDB workloads on NVMe-SSD because they provide a diverse
combination of random and sequential operations:
\begin{inparaenum}[(i)]
\item readrandom,
\item readseq,
\item readrandomwriterandom,
and
\item readreverse.
\end{inparaenum}
Class frequencies were close, suggesting that classification accuracy
is a good metric to evaluate the performance, with the least frequent
class being 21.4\% and most frequent class being 28.8\%.

We tested the neural network's performance with the aforementioned
data via k-fold cross validation with $k=10$, and found out that it
achieved an average accuracy of 95.5\%.
We also analyzed the contribution of each feature to the
classification performance; we randomized the order of a feature of
interest across samples in the validation dataset, and then calculated
the 10-fold validation performance~\cite{breiman2001random}.
Using Pearson correlation analysis~\cite{pearson1895vii}, we found that
two features were highly correlated: the cumulative moving
standard deviation and the cumulative moving mean of page offsets.
Including both would have over-emphasized their importance in
this analysis, so we excluded the cumulative standard deviation of
page offsets.
Cross validation results were 69.6\%, 76.4\%, 42.6\%, and 89.1\% for
number of transactions, cumulative moving mean of page offsets, mean
absolute page offset differences, and current readahead value,
respectively.  This shows that mean absolute page offset differences
is the most important feature, because randomizing its order reduced
the validation results the most (down to 42.6\%)---followed by number
of transactions, cumulative moving mean of page offsets, and finally
the currently used readahead value.

After obtaining classification predictions, we set the empirically
determined optimal readahead sizes according to the predicted workload
type.
In Section~\ref{sec:eval:readahead}, we evaluate the readahead
neural network not only on workloads we trained on but also
on workloads that were \emph{not} included in the training data and
workloads running on different devices (NVMe vs.\ SATA SSDs).

Figure~\ref{fig:tsne} shows that the same type of workloads for
SATA-SSD vs.\ NVMe-SSD are not placed in the same clusters all the
time.  We use neural network input data that is generated only from an
NVMe-SSD to train readahead neural network; nevertheless, we still get
significant performance improvement even for SATA-SSDs (see
Section~\ref{sec:eval:readahead}).  This indicates that our readahead
neural network is indeed learning higher-level abstractions about the
workloads, one that traditional heuristics would struggle with.

Finally, we also experimented with the readahead neural network using
TPC-H~\cite{TPC-H} queries running on MySQL~\cite{mysql20} to show how
our readahead neural network behaves on completely different types of
workloads and applications and how generalizable the models are.

\paragraph{Decision-tree models}
We also built a decision-tree (DT) model for workload type
classification based on the same features and training data.  The
readahead DT model contains 59 nodes with a maximum depth of 9.
We omit the full DT figure for brevity (available as part of our
source/data release), but for example, the decision at the root node is
whether the Z-score of the mean absolute page offset was less than
$-0.349$.
We tested the prediction accuracy of this DT using the same procedure
with the readahead neural network (10-fold cross-validation), and
observed that it results in an average prediction accuracy of only
75.4\%.
As mentioned in Section~\ref{sec:kml-stability}, \kml supports DTs
because DTs trees are more explainable than neural networks and run
considerably faster.
We evaluated the readahead DT using the same procedure as the neural
networks (Section~\ref{sec:eval:readahead}).

\paragraph{Readahead in per-file basis}
So far, we have shown how we approach the readahead problem when a
single I/O workload is accessing one device.  Storage system
developers recognize the challenge of handling mixed storage workloads
running on the same system---a common
occurrence~\cite{Amvrosiadis_2018_DSR_3316807}.
In that case, readahead values cannot be set at the device level, as
that would be suboptimal in mixed workloads.  Instead, readahead
values should be set at a higher abstraction level, on a per-file
basis.
To show our neural network's versatility, we use the same model to
tune readahead sizes not only on a per-disk basis but also on a
per-file basis.
Whereas before we ran inference every second and set one readahead
value for an entire device, here we ran inference every second on
\emph{each} open file and set a readahead value directly in Linux's
\stt{struct file}.
We evaluated the per-file basis approach and found that it could
predict and improve I/O throughput for \emph{mixed workloads} better
than both the vanilla and per-disk basis approaches (see
Section~\ref{sec:eval:readahead}).


\subsection{Use Case: NFS rsize}
\label{sec:usecase-nfs}

\paragraph{Problem definition}
Networked storage systems such as NFS are popular and heavily used.
NFS is used for storing virtual machine disks~\cite{nfs-vmdk}, hosting
NoSQL databases~\cite{linkedin-nfs}, and more.  A misconfiguration of
NFS can hurt performance.  We experimented with different applications
using NFS and found out that one critical NFS configuration parameter
is the \stt{rsize}---default network read-unit size.  Hence, we focus
predicting an optimal NFS \stt{rsize} value based on workload
characteristics.

\paragraph{Studying the problem}
We tested NFS using the same methodology as for readahead.  The only
difference here is tuning \stt{rsize} instead of readahead.  We used
NFSv4 for all of our tests.  The NFSv4 implementation we used supports
only seven different \stt{rsize} values (4K--256K).
However, in the NFS use case, there are additional external factors
not present in the readahead problem that can affect I/O performance
(\eg NFS server configuration, network speed, and number of clients
connected to the same server).
We experimented with four different RocksDB benchmarks under different
NFS server configurations and network conditions.  We configured our
server with two different NFS mount point options---one backed by
NVMe-SSD and one backed by SATA-SSD.  We injected artificial network
delays to simulate slower networks.  Our experiments revealed that
random and sequential workloads require different \stt{rsize} values
to achieve optimal performance.

\paragraph{Data collection}
We enabled NFS and page-cache related kernel tracepoints to collect
training data (\eg{} \stt{nfs4\_read}, \stt{nfs4\_readpage\_done},
\stt{vmscan\_lru\_shrink\_inactive}, and \stt{add\_to\_page\_cache}).
Unlike the readahead neural network model, we collected data from
tracepoints not only to model page cache behavior, but also network
conditions.  Similarly studying these traces, we chose our feature set
and placed our hook functions.  Our feature set includes eight
features (described below) which are calculated using the following
five data points:
\begin{inparaenum}[(i)]
\item time difference from the beginning of execution for each
  tracepoint transaction,
\item NFS file handles,
\item file offsets in NFS requests,
\item page offsets of the files that were accessed,
and
\item number of reclaimed pages during LRU scans.
\end{inparaenum}

\paragraph{Data preprocessing \& normalization}
We applied the same data preprocessing and normalization techniques
that we used for the readahead neural network.  The NFS neural network
model consists of eight features which are computed every second:
\begin{inparaenum}[(1)]
\item number of tracepoint transactions,
\item average time difference between each \stt{nfs4\_read} and
  \stt{nfs\_readpage\_done} matching pair,
\item average time difference between each consecutive \stt{nfs4\_read}
  request,
\item average time difference between each consecutive
  \stt{nfs4\_readpage\_done} request,
\item mean absolute requested offset difference between each consecutive
  \stt{nfs4\_read} request,
\item mean absolute page offset difference between each consecutive
  \stt{add\_to\_page\_cache},
\item average number of reclaimed pages,
and
\item current \stt{rsize}.
\end{inparaenum}

\paragraph{Neural network model}
We trained and tested our NFS neural network model using exactly the
same methodology as the readahead problem; for brevity, we detail only
the differences between the neural network models.
We approached the NFS problem as a workload characterization problem
and constructed our NFS neural network model with four linear layers
(with hidden layer sizes of 25, 10, and 5) with sigmoid activation
functions in between.  Similar to the readahead neural network, we
used cross entropy as the loss function and SGD as the optimizer.  We
evaluated the NFS neural network model and found out that it results
in a prediction accuracy of 98.6\% (using 10-fold cross-validation).



\section{Evaluation}
\label{sec:evaluation}

Our evaluation proceeds as follows:
First, we explain our evaluation goals in Section~\ref{sec:eval:goals}.
We then describe the testbed design and benchmarks that we used to
evaluate the readahead and NFS \stt{rsize} neural networks in
Section~\ref{sec:eval:testbed}.
In Section~\ref{sec:eval:kml} we provide performance details
regarding \kml's training and inference.
Section~\ref{sec:eval:readahead} shows how the readahead ML models
improve performance.
Finally, in Section~\ref{sec:eval:nfs}, we present our evaluation of the
\stt{rsize} neural network model for NFS.

\subsection{Evaluation Goals}
\label{sec:eval:goals}

Our primary evaluation goal is to show that using ML techniques inside
the OS can be used to to tune parameters dynamically and improve
storage systems' performance.

We start by showing the practicality of using ML in kernel space.
We evaluate \kml's system overheads in terms of (i) data collection
overhead, (ii) training cost, (iii) inference cost, and (iv) memory
usage.
Then, we evaluate both readahead and NFS neural network models to show
how they improve the I/O performance and quickly adapt the system in
the presence of changing workloads and conditions.
To show that our models can learn
abstract workload patterns, we first present the generalization power
of our models by testing it on workloads \emph{not} included in the
training dataset.
Next, we present benchmarks on a device type that was \emph{not} used
in the data collection phase or training.
We also built a decision tree model for the readahead problem to have
\emph{comparable} results since decision trees are more explainable,
still popular, and closer in operation to traditional heuristics.

Furthermore, we evaluate \kml's versatility by applying the readahead
neural network model on a per-file basis.
This demonstrates \kml's ability to optimize individual I/Os in a
mixed workload.
Lastly, we evaluate our readahead ML models' behavior when they
mispredict and how quickly they recover.


\subsection{Testbed}
\label{sec:eval:testbed}

We ran the benchmarks on two identical Dell R-710 servers, each with
two Intel Xeon
quad-core CPUs (2.4GHz, 8 hyper-threads), 24GB of RAM
and an Intel 10GbE NIC.
In some experiments, we intentionally configured the system with only
1GB of memory to force more memory pressure on the I/O system; but we
also show experiments with the full 24GB of system RAM.  We used the
CentOS 7.6 Linux distribution.
We developed \kml for Linux kernel version 4.19.51, the long-term
support stable kernel;
we added our readahead ML models
to this kernel and used it in all experiments.
Because HDDs are becoming less popular in servers, especially when I/O
performance is a concern, we focused all of our experiments on SATA
and NVMe SSDs.
We used Intel SSDSC2BA200G3 200GB as our SATA-SSD device and a Samsung
MZ1LV960HCJH-000MU 960GB as our NVMe-SSD device, both formatted with
Ext4.
These two devices were used exclusively for RocksDB databases.
To
avoid interference with the installed CentOS,
the two servers have a dedicated Seagate ST9146852SS 148GB SAS boot
drive for CentOS, utilities, and RocksDB benchmark software.
We used 10GbE switches to connect the machines (useful for NFS
experiments).
We observed an average RTT time of 0.2 milliseconds.

\paragraph{Benchmarks}
We chose RocksDB's \stt{db\_bench} tool to generate diverse workloads
for evaluating the readahead
and NFS \stt{rsize}
neural networks.
RocksDB~\cite{rocksdb} is a popular key-value store and covers an
important segment
of realistic storage systems; \stt{db\_bench} is a versatile
benchmarking tool that includes a diverse set of realistic workloads.
Workloads can be run individually or in series, and the working set
size can be easily configured to generate more I/O
pressure on a system.
On the 1GB RAM systems, we configured a RocksDB of twice the size
(2GB).

To demonstrate that our ML models can learn from and optimize for
different types of real-world workloads, we chose the following six
popular yet different \stt{db\_bench} workloads:
\begin{inparaenum}[(1)]
\item readrandom,
\item readseq,
\item readrandomwriterandom (alternating random reads and writes),
\item readreverse,
\item updaterandom (read-modify-write in random offsets),
and
\item \stt{mixgraph} (a complex mix of sequential and random accesses,
  based on Facebook's realistic data that follow certain Pareto and
  power-law distributions~\cite{cao2020characterizing}).
\end{inparaenum}

We trained our readahead neural network on traces that contain only
four of these workloads: readrandom, readseq, readreverse,
readrandomwriterandom---all running only on the NVMe-SSD.
These four tend to be the simpler workloads, because we wanted to see
whether \kml can train on simpler workloads yet accurately predict on
more complex workloads not trained on.
This also ensures a balanced representation of randomness and
sequentiality in the training dataset.

After the training phase completed, we tested our models on all six
workloads as well as different devices.  This was done to show that our
models not only perform accurate predictions on
the training set samples, but they also generalize
to two new and complex workloads (updaterandom and
\stt{mixgraph}
as well as a different device (SATA-SSD))---which were
excluded from the training data.
We evaluated mixed workloads by running two concurrent \stt{db\_bench}
instances, each on a separate RocksDB database and using a different
workload profile, both stored on the same device.
We kept the hardware configuration the same as before (1GB RAM)
to increase system and page-cache pressure.

We also experimented with our readahead network model using
TPC-H~\cite{TPC-H} queries running on MySQL~\cite{mysql20}, to
evaluate how generalizable and effective the readahead neural network
is to an entirely different workload.
In this paper we do \emph{not} claim that our readahead neural network
model will work universally to optimize readahead values for all
possible workloads.  Rather, these use cases are meant to demonstrate
the \kml framework's versatility.  With more workloads and datasets,
one can build a wide range of ML models to optimize many storage
problems.


\subsection{\kml's Overheads}
\label{sec:eval:kml}

An ML model's overhead depends on its architecture.  Generally, deeper
or higher-dimensional neural networks consume more memory and CPU
than, say, decision-tree models.  It is vital that an ML component,
especially one that may run inside the kernel, consume as little CPU
and memory as possible.  Next, we evaluate the readahead neural
network overheads.

\begin{figure}[t]
  \centering
  \includegraphics[width=1.0\columnwidth]{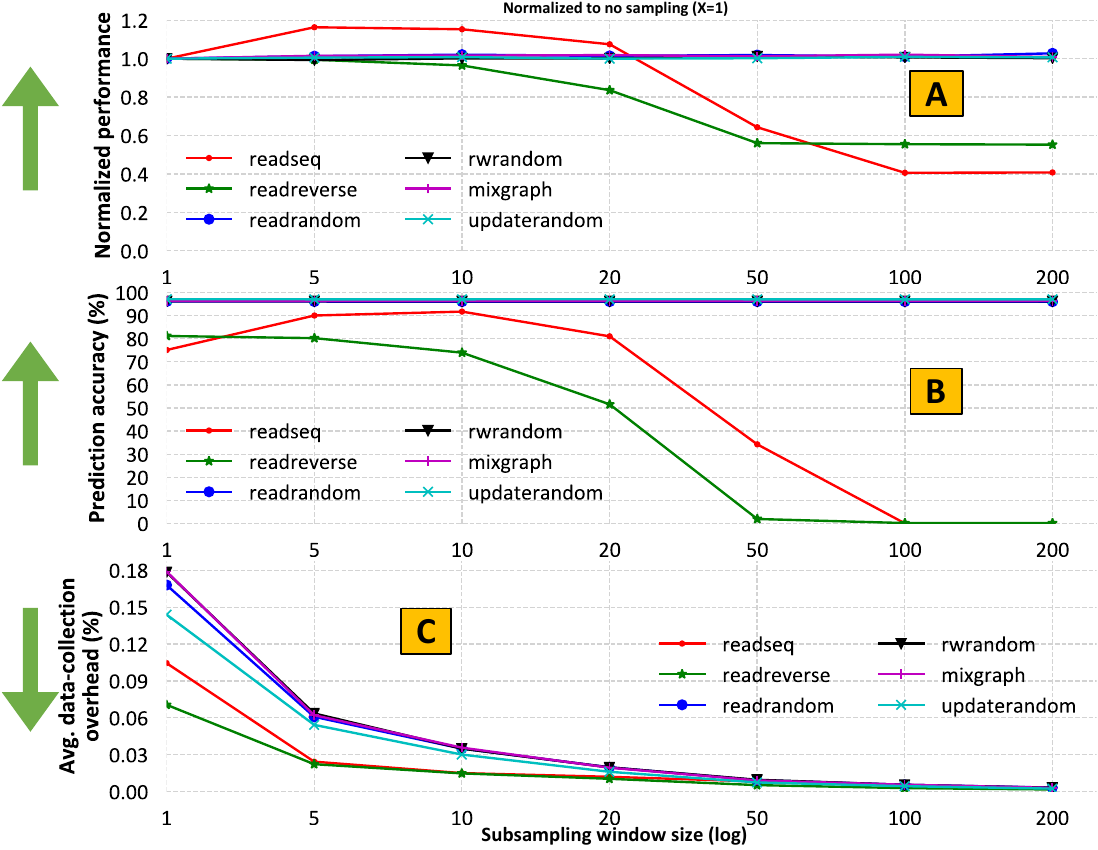}
  \caption{Performance (A), prediction accuracy (B) and CPU overheads
    (C) in seven different subsampling window sizes for the per-disk
    readahead neural network.  Upward green arrows denote that higher
    is better.}
  \label{fig:subsampling-overall-eval}
\end{figure}




\begin{figure*}[thbp]
  \centering
  \includegraphics[width=1.0\textwidth]{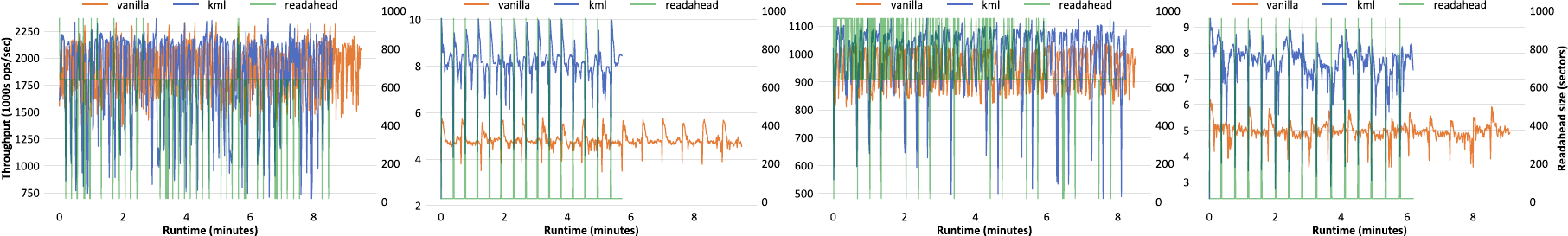}
  \caption{Running four back-to-back RocksDB workloads in order from
    left to right: readsequential, readrandom, readreverse, then
    mixgraph.  Here, we started with the default readahead value;
    thereafter, the last value set in one workload was the one used in
    the next run.  For each of the four graphs, we show their Y axes
    (throughput, different scales).  The readahead value is shown as
    the Y2 axis for the rightmost graph (d) and is common for all
    four.  Each workload ran 15--50 times in a row, to ensure we ran
    it long enough to observe patterns of mis/prediction and reach
    steady-state.  Again, we see \kml adapting, picking optimal
    readahead values, occasionally mis-predicting but quickly
    recovering, hence overall throughput was better.}
  \label{fig:back-to-back}
\end{figure*}


\paragraph{Data gathering overheads}
The only inline operations that readahead neural network inserts
directly in the data path are data collection probes.
Hence it is vital for these probes to be optimized.
Figure~\ref{fig:subsampling-overall-eval}(C) shows how the data
collection CPU overheads (percentage) change with subsampling window
sizes.  When there is no subsampling in the system ($X=1$), the CPU
overheads of data collection probes is as high as 0.18\%.  Although
this is a fairly low overhead considering the multiplicative I/O
benefits we report, this overhead can be reduced further by increasing
the subsampling window.
However, increasing the subsampling windows size can hurt prediction
accuracy and performance improvements, as less data is available to
make rapid predictions.  see
Figure~\ref{fig:subsampling-overall-eval}(A) and~(B).
Figure~\ref{fig:subsampling-overall-eval}(B) shows that workloads with
a lot of randomness in them were the least affected, because
randomness is still predicted as random even with fewer samples; yet
we can reduce the already small CPU overheads even more.

The figure further shows that only sequential workloads are affected
by subsampling window changes: generally, as the sampling window
widens, prediction accuracy and normalized performance worsen.
However, we noticed an unexpected behavior for the \stt{readseq}
workload.  Increasing the subsampling window size from one to five or
ten \emph{actually improved} both prediction accuracy and performance;
this is because \stt{readseq} keeps the I/O subsystem busy at near
maximum bandwidth, and increasing subsampling window size reduced
short-term noise that resulted in more frequent mispredictions.

We can also observe that the data collection overheads depend on the
workload type.  For example, \stt{readseq} workload's average data
sampling frequency per-second is around 30K but its data collection
overhead is still lower than \stt{mixgraph} workload which has 20K
average data sampling frequency per-second.  The reason that data
collection overheads change based on the workload type, is due to the
sudden I/O bursts resulting in some cache misses (bi-modal histograms
omitted for brevity).
%

\paragraph{Inference/training overheads}
The readahead neural network performs inference (prediction) and
changes the block-layer readahead value in 21$\mu$s on average
(std.\ dev.\ $<10\%$).  This action executes in a separate,
asynchronous kernel thread, once in every second.  Hence, it has
negligible impact on the overall OS performance.
When the readahead neural network runs in per-file mode, \kml runs
inferences an average 135 times a second (\ie one per open file):
inferencing for all open files consumes 1.7ms on average.
We measured that the readahead decision tree inference takes only
8$\mu$s (using the same feature vector).
The readahead neural network and decision tree have the same
data pre-processing and normalization implementation---the only
difference between them is in the inference part.
Overall, these overheads are fairly small and acceptable, considering
the multiplicative I/O performance benefits they enable.

As discussed in Section~\ref{sec:usecase-readahead}, our readahead
neural network prototype offloads training to the user level.  We
measured the time to perform one training iteration in user level at
51$\mu$s on average; this training iteration includes the forward
pass, back-propagation, and weight update stages.

\paragraph{Memory overheads}
The readahead neural network allocates 3,916 bytes of dynamic memory
during the model's initialization phase.  While inferencing, \kml
temporarily allocates 676 bytes before returning the inference
results.  This overall memory footprint is negligible in today's
multi-GB systems.  The readahead decision tree occupies only 2,432
bytes of dynamic memory during initialization.  The decision tree
model does not allocate dynamic memory during inference.
Lastly, the kernel module \stt{readhead.ko} has a binary memory
footprint of 432KB and the kernel module \stt{nfs.ko} is 636KB, while
the \kml framework itself (\stt{k-Mlib.ko}) has a memory footprint of
5.5MB.

\paragraph{Practicality and scalability}
Our vision is \kml could enable a future where traditional heuristics
are gradually replaced with ML-based approaches to improve storage and
network I/O performance.
In Section~\ref{sec:eval:readahead} we demonstrate, for example, that
our readahead neural network model improves I/O performance by as much
as 2.3$\times$, but consumes less than 0.2\% additional CPU cycles: we
believe this is a fairly acceptable trade-off for most users.
Nevertheless, we tested this model with 100
concurrent inferences and found that both overheads
and I/O improvements have scaled linearly; hence \kml's benefits still
outweigh its overheads.


\subsection{Readahead Evaluation}
\label{sec:eval:readahead}

%
\begin{figure*}[htbp]
  \centering
  \includegraphics[width=1.0\textwidth]{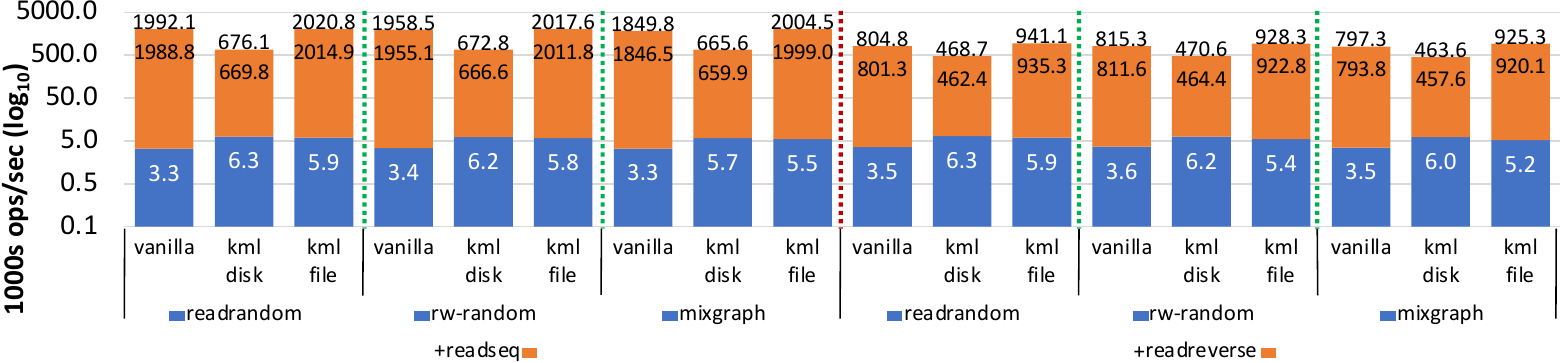}
  \caption{Mixed workloads results.  We ran sequential and random
    workload combinations on the same NVMe-SSD device.  Each unique
    combination is tested with the readahead neural network running in
    per-disk basis (kml disk) and per-file basis (kml file) and
    compared against vanilla results.  The model running in per-file
    basis outperformed both vanilla and per-disk modes.}
  \label{fig:ultimate}
\end{figure*}


%
\begin{figure}[thbp]
  \centering
  \includegraphics[width=1.0\columnwidth]{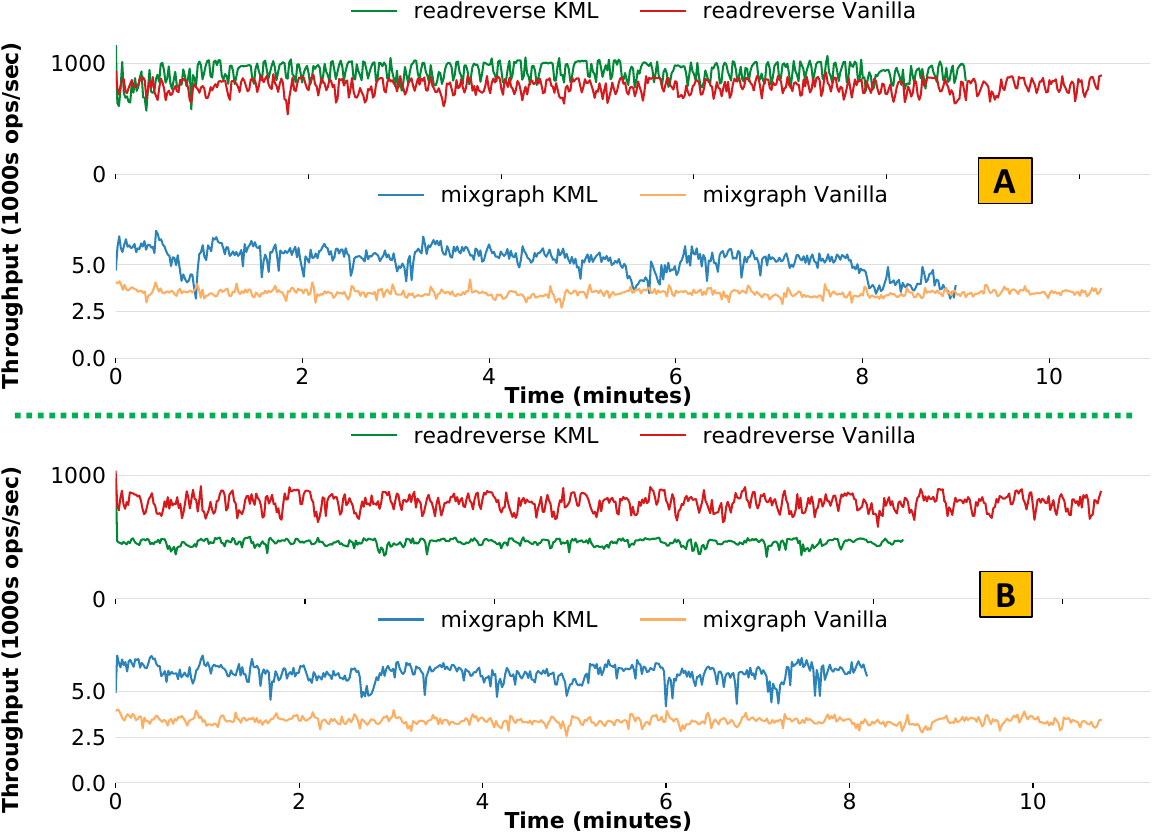}
  \caption{Mixed workloads results on a timeline, comparing the
    readahead neural network model running on per-file basis ('A',
    top) vs.\ per-disk basis ('B', bottom).}
  \label{fig:per-disk-vs-per-file}
\end{figure}




\paragraph{Readahead background}
There are two places in the Linux kernel where readahead is defined:
the block layer and the file system level.  When a file is opened, the
VFS initializes an open \stt{struct file} and copies the readahead
value for that file from the corresponding block layer.  Upon a page
fault for that file, the page-cache layer uses the value stored in the
file to initiate reading-ahead the desired number of sectors of that
file.
However, the readahead value in the file structure is initialized only
once when the file is opened.  So when \kml changes the block layer
readahead value, the Linux kernel does not copy the new value to any
file already opened.  This means that open files may continue to use a
sub-optimal readahead value, even if better values are available (\eg
due to workload changes).
That is why we implemented a mechanism that changes the readahead size
for \emph{open files} when \kml changes the disk-level readahead
value.  This propagates newer readahead values to each open file,
improving our adaptability.
Conversely, if \kml mispredicts the workload type and changes the
readahead size to a suboptimal value, short-term performance
degradation can happen, which might hurt overall performance.

\paragraph{Back-to-back workloads on NVMe}
Figure~\ref{fig:back-to-back} shows four workloads running back to
back with each subfigure comparing a vanilla run (colored orange) to
our \kml-enabled readahead run (colored blue).  The readahead value
was left at the default value (\ie 256) at the start of both vanilla
and \kml-enabled runs, but when the next workload started, it used the
last readahead value from the previous workload's run (\eg the
readahead value at the end of the leftmost subfigure is the same at
the start of the subfigure immediately to its right).
This experiment evaluates \kml's ability to optimize the readahead
values when the I/O workload may change every few minutes.
The X axes indicate the run time in minutes.  The Y axes indicate
throughput in thousands of ops/sec (higher is better), and have
different scales for each experiment.
The Y2 axes show the readahead values used or predicted by
\kml over time in terms of number of sectors (denoted with a green line
and using the same scale).
Each workload ran 15--50 times in a row, so it ran long enough to
observe mis/predictions patterns.
As seen in Figure \ref{fig:back-to-back}, \kml adapts quickly to
changing workloads by tuning the readahead value in about one second.

Although we observe some mis/prediction patterns, seen as sudden
spikes, overall throughput still improved across all four runs,
averaging 63.25\% improvement: 140\% improvement for readrandom, 2\%
for readsequential, 109\% for \stt{mixgraph}, and 12\% for
readreverse.
We note that even a small improvement in throughput can yield
significant cumulative energy and economic cost savings for
long-running servers~\cite{li-hotstorage2014-cost}.

\begin{figure}[t]
  \centering
  \includegraphics[width=1.0\columnwidth]{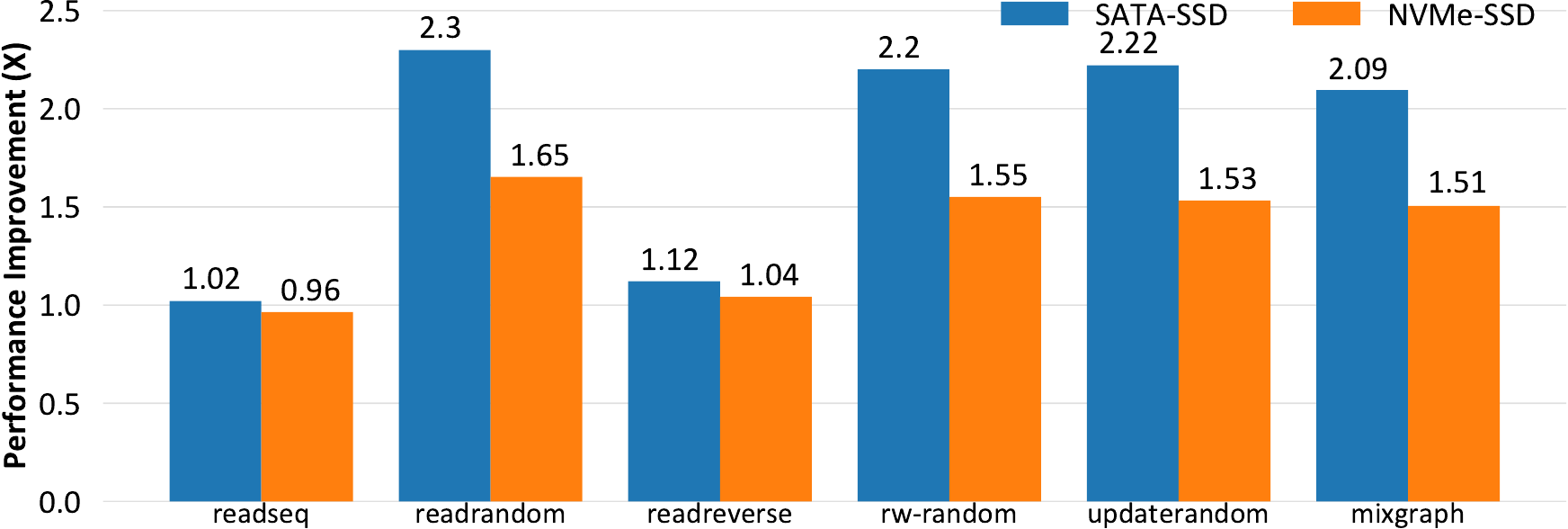}
  \caption{Readahead neural network performance improvements
    ($\times$) for RocksDB benchmarks on SATA-SSD and NVMe-SSD across
    all six workloads, normalized to vanilla (1.0$\times$).}
  \label{fig:overall-performance}
\end{figure}

\paragraph{Read-sequential workloads}
Out of the six workloads we ran, Figure~\ref{fig:overall-performance}
shows the one where \kml performed the worst: read-sequential.
Reading data sequentially directly from the raw SATA-SSD is nearly
1,000$\times$ faster than the \stt{mixgraph} workload, and nearly
400$\times$ faster with the NVMe-SSD.
Here, there is little opportunity for \kml to improve throughput for a
sequential workload that reads at speeds near the maximum throughput
of the physical device.
%

\paragraph{Read-reverse workloads}
As we can see from the fluctuating green line (readahead values in
Figure~\ref{fig:back-to-back}) \kml mispredicts readreverse as
readseq
and changes the readahead value to something suboptimal.  These two
workloads both access files sequentially---one reading forward and one
backward.  Interestingly, readseq and readreverse are quite close from
a feature representation perspective, which explains the
mispredictions.  But since both of these workloads access files
sequentially, their optimal readahead values are also quite close to
each other.  Thus, even when \kml mispredicts readreverse as readseq
or vice versa, this had a small overall impact on performance.

\paragraph{Summary of readahead neural network results}
We summarize all readahead neural network results in
Figure~\ref{fig:overall-performance}.  We observe that the average
throughput improvement for NVMe-SSD is ranging from 0\% to 65\%.  We
saw greater improvements in the SATA-SSD case, ranging from 2\% to
130\% (2.3$\times$).
Lastly, we ran the complex \stt{mixgraph} workload on NVMe-SSD with
the system memory set to the maximum (\ie 24GB) and the database size
set to be relatively large, 65GB (compared to a 2GB baselines database
size).
This experiment ran for nearly an hour (48.5 minutes) and resulted in
an average throughput improvement of 38\%.

\paragraph{Mixed workloads}
Mixed workloads are considered a challenging optimization
problem~\cite{Amvrosiadis_2018_DSR_3316807}.
In Figure~\ref{fig:per-disk-vs-per-file}, we present a timeline
performance comparison using the readahead neural network model
running on a per-disk vs.\ per-file basis.
The per-file mode performs better overall because readahead values are
set for each open file independently.
Conversely, in the per-disk mode, a single readahead value is set at
the disk level and hence uniformly on all open files: a readahead
value good for one workload is likely to be suboptimal for other open
files.
One reason why the per-disk mode cannot predict workload types
correctly is that when different workloads are mixed---even sequential
ones or ones with regular patterns---the mix looks more like a purely
random workload at the disk level.
Figure~\ref{fig:ultimate} shows overall mixed workloads performance
comparisons.
Per-file mode performed overall better in every combination of mixed
workloads.
If we compare only the sequential parts of the mixed workload
combination (orange bars in Figure~\ref{fig:ultimate}), in per-disk
mode, we observe significant performance degradation.
However, in per-file mode, we can observe performance improvements for
both the sequential and random (blue bars in
Figure~\ref{fig:ultimate}) parts of the mixed workload combination.
The reason why per-disk mode performs better for the random parts of
the mixed workload combinations is for the same reason: mixing
workloads looks more random-like at the disk level.  \kml predicts
these as readrandom or readrandomwriterandom which coincidentally fits
this part of the workload, but significantly hurts non-random
workloads.

\paragraph{Decision tree evaluation}
\begin{figure}[t]
  \centering
  \includegraphics[width=1.0\columnwidth]{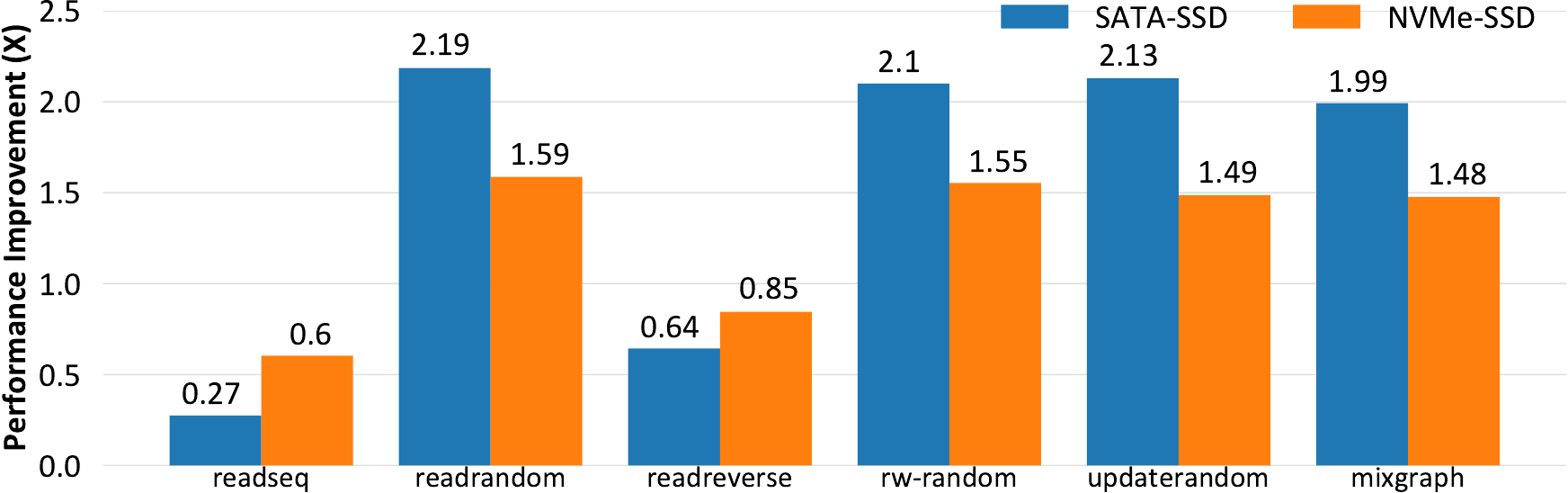}
  \caption{Readahead decision tree performance improvements ($\times$)
    for RocksDB benchmarks on SATA-SSD and NVMe-SSD devices across all
    six workloads, normalized to vanilla (1.0$\times$).}
  \label{fig:dt-overall-performance}
\end{figure}


%
\begin{figure}[t]
  \centering
  \includegraphics[width=1.0\columnwidth]{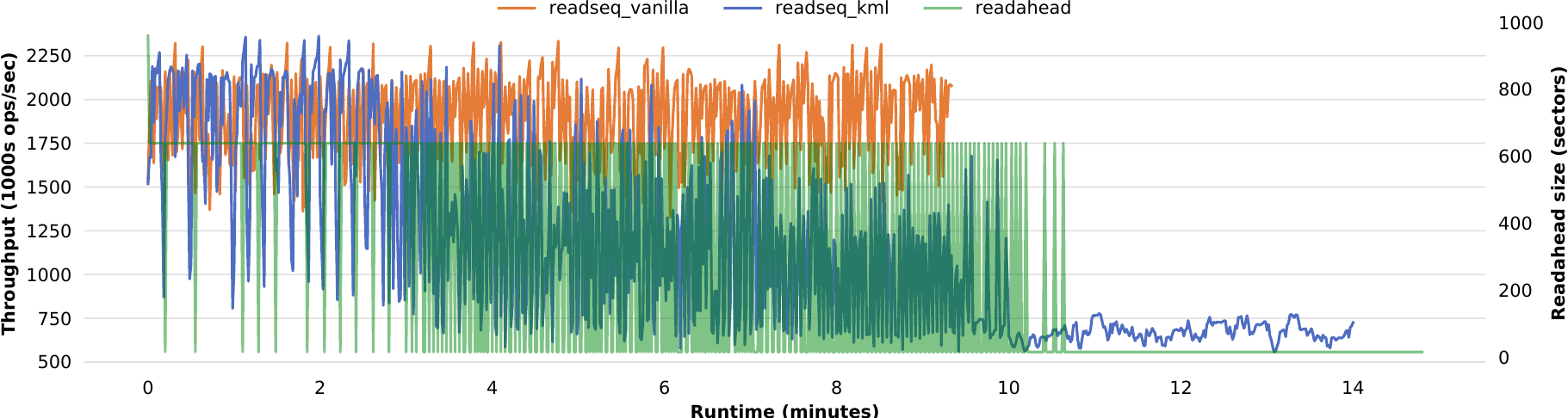}
  \caption{Performance timeline graph for tuning with \kml decision
    tree while running readseq workload on NVMe-SSD.}
  \label{fig:dt-readseq-performance}
\end{figure}


%
In addition to the neural network model, we implemented a decision
tree model for the readahead problem to compare the two ML approaches
on the same problem.  We tested the readahead decision tree the same
way.  Figure~\ref{fig:dt-overall-performance} shows that there is a
performance improvement for workloads with a random component.
For the readahead decision tree, we measure average throughput
improvement for random workloads on NVMe-SSD as ranging from 48\% to
59\%; and in the SATA-SSD case, ranging from 99\% to 119\%
(2.19$\times$).  While good, the neural network model yielded greater
improvements, as discussed above.

The DT model, however, degraded performance for sequential workloads.
it degraded performance for sequential workloads on NVMe-SSD by
15--40\%; and in the SATA-SSD case, by 36--73\% worse.
We investigated this performance degradation.
Figure~\ref{fig:dt-readseq-performance} shows the readseq workload
running on a RocksDB instance stored on an NVMe-SSD.  Here, the
readahead decision tree predicts the workload correctly in the first
three minutes, despite some fluctuations.  Afterwards, the decision
tree model's predictions fluctuate wildly, and at around minute 10 it
consistently makes wrong predictions.
Overall, this was somewhat expected for our I/O optimization problem:
neural network models, while more complex to train and use, are more
adaptable than decision-trees~\cite{hall2003neural}.  Specifically,
when the DT model mispredicts, and system conditions change (\ie I/O
activity), the DT model continues to mispredict, and it cannot recover
as quickly as the more adaptable neural network model.

\paragraph{TPC-H benchmarks}
\begin{figure}[t]
  \centering
  \includegraphics[width=1.0\columnwidth]{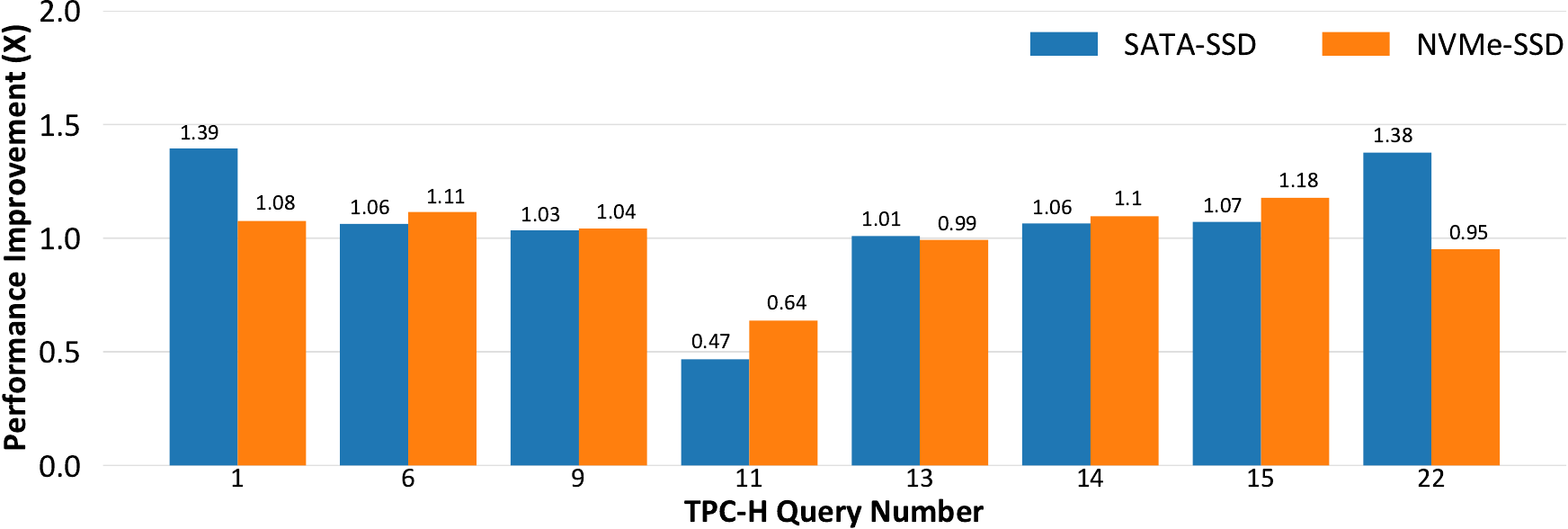}
  \caption{Readahead neural network performance improvements
    ($\times$) for TPC-H queries on SATA-SSD and NVMe-SSD devices,
    normalized to vanilla (1.0$\times$).}
  \label{fig:tpch-overall-performance}
\end{figure}


%
As we mentioned in Section~\ref{sec:eval:testbed}, we evaluated our
readahead neural network model---trained on RocksDB
workloads---on TPC-H queries (both with NVMe-SSD).
This intends to show the model's accuracy limitations when presented
with vastly different workload and application combinations.
Figure~\ref{fig:tpch-overall-performance} shows performance
improvements as much as 39\% for most query types.
For query 11, however, the readahead neural network failed to
characterize the workload correctly and resulted in a 53\% performance
reduction.
Nevertheless, overall TPC-H performance still improved by 6\%.
We expect that neural network models trained on more traditional SQL database
workloads would likely yield even better predictions across most similar databases.


\subsection{NFS Evaluation}
\label{sec:eval:nfs}
\begin{figure}[t]
  \centering
  \includegraphics[width=1.0\columnwidth]{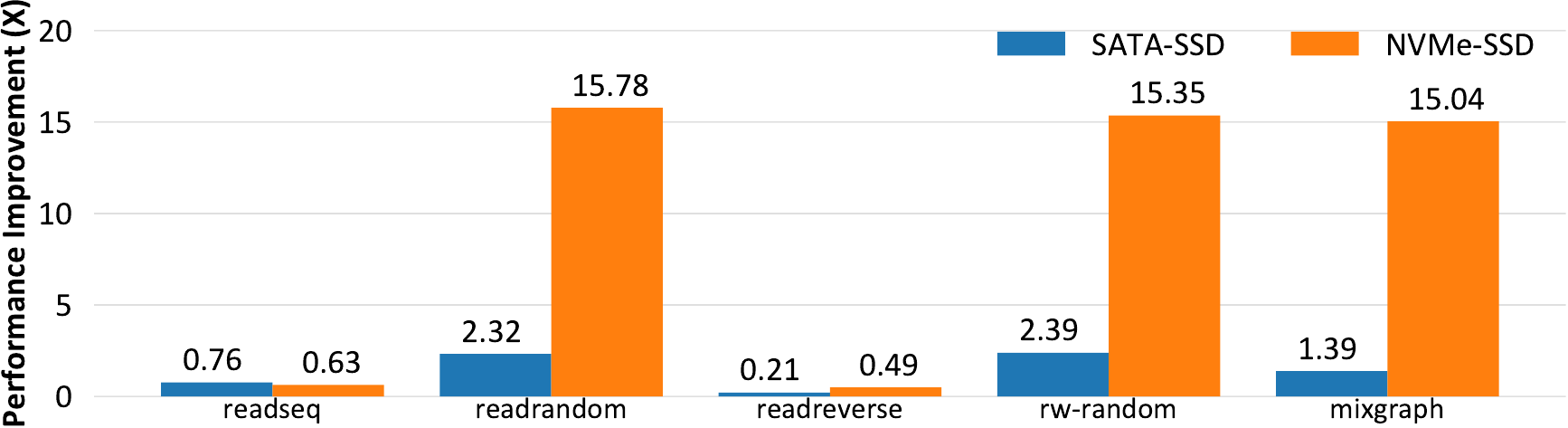}
  \caption{Performance improvements ($\times$) for RocksDB benchmarks
    on SATA-SSD and NVMe-SSD devices across all six workloads running
    on NFS, normalized to vanilla (1.0$\times$).}
  \label{fig:nfs-overall-performance}
\end{figure}


%
Figure~\ref{fig:nfs-overall-performance} shows the NFS \stt{rsize}
neural network performance improvements using the same evaluation
techniques of readahead.  Throughout these experiments, we ran
multiple iterations of the same workloads.
Since \stt{rsize} is a mount point parameter for NFS, our NFS neural
network can tune \stt{rsize} values only in the beginning of the
iteration.  (We plan to fix the Linux kernel to permit \stt{rsize} to
change dynamically.)  Hence, in sequential workloads, if the NFS
neural network makes even one misprediction, it will affect the entire
iteration, leading to performance degradation.  Nevertheless, in
random workload cases, we still measured around 15$\times$ performance
improvement; in separate experiments (not shown for brevity),
performance improvements for random workloads reached up to
20$\times$.  This demonstrates the significant potential of \kml.



\section{Related Work}
\label{sec:related}

\paragraph{Machine learning in systems and storage}
In follow-up work to Mittos~\cite{hao2017supporting}, a custom neural
network was built that makes inferences inside the OS's I/O scheduler
queue.  The neural network decides synchronously whether to submit
requests to the device using binary
classification~\cite{hao2020linnos}.  There are notable differences
between that system and our \kml.  That system was trained offline
using TensorFlow and exclusively trained in user space.
Additionally, each of their two layers were custom built.
Conversely, \kml provides a more flexible architecture.  \kml
training, retraining, normalization, repeated inference---all are
possible and accomplished with ease in any combination of online,
offline, synchronous, or asynchronous settings.  Lastly, \kml easily
supports an arbitrary number of generalizable neural network layers;
our experiments demonstrate more expressive classification abilities
on a more diverse set of devices.

Laga \etal~\cite{laga2016lynx} improved readahead performance in the
Linux Kernel with Markov chain models, netting a 50\% I/O performance
improvement in TPC-H~\cite{TPC-H} queries on SATA-SSDs.
In contrast, our experiments ran on a wider selection of storage media
(NVMe-SSD and SATA-SSD) and workloads.  In TPC-H, we show improvements
up to 39\% despite TPC-H being a completely new workload for our
readahead model.  Moreover, our results illustrate that our readahead
model can improve I/O throughput by as much as 2.4$\times$---all while
keeping memory consumption under 4KB, in comparison to Laga \etal's
much larger 94MB Markov chain model.

Parameter tuning for storage and operating systems has been a
challenge and researchers approached this problem using control
theory~\cite{sironi2012metronome} and data distribution analysis for
storage clusters~\cite{abd2005ursa}.
Some research has attempted to apply ML techniques to OS task
scheduling~\cite{negi2005applying, chen2020mlloadbalancing}, with
small reported performance improvements (0.1--6\%).
Nevertheless, it is becoming increasingly popular to apply ML
techniques to storage and OS problems including: tuning SSD
configurations~\cite{li2021learning}, memory
allocation~\cite{maas2020learning}, TCP congestion~\cite{dong2018pcc},
building smart NICs~\cite{siracusano2020running}, predicting index
structures in key-value stores~\cite{kraska2018case, dai2020wisckey},
offline black-box storage parameter
optimization~\cite{zhen2018towards}, reconfigurable kernel datapaths~\cite{qiu2021toward},
local and distributed
caching~\cite{vietri-hotstorage18-lecar, subedi2018stacker}, database
query optimization~\cite{sagedbkraska}, and cloud resource
management~\cite{cortez2017resource, delimitrou2013paragon,
  delimitrou2014quasar, somashekar2021towards}.

\paragraph{Machine learning libraries for resource-constraint systems}
A myriad of ML libraries exist---some general purpose and others more
specialized.  Popular general-purpose ML libraries include
Tensorflow~\cite{abadi2016tensorflow},
PyTorch~\cite{paszke2019pytorch}, and CNTK~\cite{cntk}.
Conversely, libraries like ELL~\cite{microsoft-ell}, Tensorflow
Lite~\cite{tensorflow-lite}, SOD~\cite{SOD}, and Dlib~\cite{Dlib}
specialize to run on resource-constrained or on-device environments,
\kml differentiates itself by targeting OS-level applications and
designed for OS and storage systems specifically.  Inside the OS,
resources are \emph{highly} constrained, prediction accuracy is vital,
and even small data-path overheads are unacceptable.

\paragraph{Adapting readahead and prefetching}
Readahead and prefetching methods are both well-studied
problems~\cite{Shriver1998AnAB, shriver1999does, ding2007diskseen,
  kroeger2001design} and see use in distributed
systems~\cite{lee2018aps, tran2004automatic, chen2019rnn,
  dong2010correlation, liao2015performing, nalajala2019improving,
  liang2007step, cherubini-icdm17-prefetching}.
Many have attempted to build statistical models to optimize and tune
systems~\cite{Shriver1998AnAB, shriver1999does, fox2008quantifying}.
However, the main limitation of statistical models is their inability
to adapt to novel new workloads and devices.  We have shown that our
model can adapt to \emph{never-before-seen} workloads and devices.
Another way to improve a readahead system is to predict individual I/O
requests and file accesses by observing workload
patterns~\cite{ding2007diskseen, kroeger2001design, Wu2008OnTD,
  uppal2012flashy, hu2010feature, amer2002file, whittle2003using,
  xu2020frequent}.
Predicting file accesses using hand-crafted algorithms is a reasonable
first approach.  However, such manual labor simply cannot keep up with
the diversity and complexity of ever-changing modern workloads.
Conversely, as long as we have training data, ML models can adapt,
retrained as needed, and optimize much faster.
Simulations are also viable solutions for readahead and
prefetching problems~\cite{ganfure2020deepprefetcher,
  chakraborttii2020learning, ravichandran2005making, xu2011evaluation,
  zheng2017adaptive}.
However, simulations are computationally expensive and are limited to
the datasets that the models are trained and tested with.
Additionally, the models produced in simulations are not designed for
resource-constrained environments, making it non-trivial to migrate
such models to the kernel.  It is possible to use a user-space library
to intercept file accesses~\cite{won2018ifetcher} or to require
application-level changes~\cite{yang2002atc}.  In contrast, \kml
requires no application changes and is capable of intercepting
\stt{mmap}-based file accesses.

Finally, while techniques exist to improve NFS performance, we are
unaware of automated ones that use ML~\cite{improvenfswrite}.


\section{Conclusion}
\label{sec:conclusion}

Operating systems and storage systems have to support many
ever-changing workloads and devices.  To provide the best performance,
we have to configure storage system knobs based on workloads' needs
and device characteristics.  Unfortunately, current heuristics cannot
adapt to workload changes quickly enough and require constant
development efforts to support new devices.  We propose \kml to solve
these problems---an ML framework inside the OS that adapts quickly to
optimize storage performance.  \kml enables finer granularity
optimizations for individual files in even mixed workloads---a
challenging problem.  Our preliminary results show that, for a
readahead problem, we can boost I/O throughput by up to 2.3$\times$
without imposing significant CPU/memory overheads.  For the NFS
\stt{rsize} problem, the improvement was up to 15$\times$.  These I/O
throughput improvements far outweigh the small memory and CPU
consumption of \kml.

\paragraph{Future work}
We plan on using \kml to tune knobs for other OS subsystems: \eg
packet and I/O schedulers, and networking.  We are
adding  ML techniques to \kml, such
as reinforcement learning~\cite{kaelbling1996reinforcement}, which can
be a better fit for solving certain OS problems.
To support more advanced ML approaches (\eg Recurrent
Neural Networks (RNNs)~\cite{wiki-rnn}) and Long Short-Term Memory
(LSTM)~\cite{hochreiter1997long}), we are extending \kml to support
arbitrary computation DAGs.
We also plan to integrate user-kernel co-operated design into \kml.
Finally, loading an unverified ML model into a running kernel opens up
new attack surfaces.  We are exploring known techniques to digitally
sign and certify loadable models~\cite{linux-code-signing,
  kim2018broken}.


\section{Acknowledgments}
\label{sec:acks}

This work was made possible in part thanks to Dell-EMC, NetApp, and
IBM support; and NSF awards CCF-1918225, CNS-1900706, CNS-1729939, and
CNS-1730726.


\bibliographystyle{plain}
\bibliography{master,local}

\end{document}
\fi
